\documentclass[runningheads]{llncs}
\usepackage[T1]{fontenc}
\usepackage{graphicx}
\usepackage{makecell, multirow, lineno, hyperref}
\usepackage{booktabs}
\usepackage{amsmath}
\usepackage{utfsym}

\begin{document}%
\title{Improving Target Speaker Extraction with Sparse LDA-transformed Speaker Embeddings}

\author{Kai Liu \and Xucheng Wan \and Ziqing Du
\and Huan Zhou}
%

\institute{AI Application Research Center, Huawei Technologies Co., Ltd. Shenzhen, China \\
\email{\{liukai89,wanxucheng,duziqing1,zhou.huan\}@huawei.com}}
\maketitle              
\begin{abstract}
As a practical alternative of speech separation, target speaker extraction (TSE) aims to extract the speech from the desired speaker using additional speaker cue extracted from the speaker. Its main challenge lies in how to properly extract and leverage the speaker cue to benefit the extracted speech quality. The cue extraction method adopted in majority existing TSE studies is to directly utilize discriminative speaker embedding, which is extracted from the pre-trained models for speaker verification. 
Although the high speaker discriminability is a most desirable property for speaker verification task, we argue that it may be too sophisticated for TSE. 
In this study, we propose that a simplified speaker cue with clear class separability might be preferred for TSE. To verify our proposal, we introduce several forms of speaker cues, including naive speaker embedding (such as, x-vector and xi-vector) and new speaker embeddings produced from sparse LDA-transform. Corresponding TSE models are built by integrating these speaker cues with SepFormer (one SOTA speech separation model). Performances of these TSE models are examined on the benchmark WSJ0-2mix dataset. Experimental results validate the effectiveness and generalizability of our proposal, showing up to 9.9\% relative improvement in SI-SDRi. Moreover, with SI-SDRi of 19.4 dB and PESQ of 3.78, our best TSE system significantly outperforms the current SOTA systems and offers the top TSE results reported till date on the WSJ0-2mix.

\end{abstract}

\section{Introduction}
In real multi-talker communications, like human conversations or multi-party meetings, it is a common scenario that the speech from a speaker of interest is interfered with others. Such a corrupted speech mixture may bother human listeners with partially intelligible speech or make adverse impacts for downstream applications (such as speech recognition and speaker diarization). To get the desired speech from such a complex auditory scene, is a longstanding challenge in the field of computational auditory scene analysis \cite{wangbook06}. With the advent of deep neural network (DNN), two DNN-based approaches, speech separation (SS) and target speaker extraction (TSE) algorithms have been proposed to solve the problem with distinctive objectives.

The objective of SS algorithms is to recover all individual speakers from the mixture signal. Given target, mixture signal and the overall speaker number, SS can be formulated as a supervised learning problem. Most SS algorithms estimate time-frequency separation masks or filters that are optimized to minimize the signal reconstruction loss. With large amount of training data, such an approach can deliver impressive separation performance. On the most widely used dataset WSJ0-2mix \cite{hershey16}, it is reported that the SepFormer \cite{subakan21}, as one of the top SS models, achieves performance that is close to the upper bound of the dataset (22.3 vs. 23.1 dB), in terms of scale-invariant signal-to-distortion ratio improvement (SI-SDRi) \cite{lutati22}. 

Alternatively, TSE aims to extract the speech from the target speaker while ignoring others, making it more practical for real deployment. As a special case of SS, TSE outputs one single speech and uses additional speaker identification information (called speaker cue herein) as auxiliary input. 

Many forms of auxiliary input are available and the most common one is an pre-recorded enrollment utterance from target speaker. However, it is non-trivial to extend a well-performed SS model to a TSE one that still maintains high performance. The underlying challenge mainly lies in how to properly extract and leverage the speaker cue to benefit the extracted speech quality. In this paper, we are interested in the challenge on proper extraction of speaker cue that captures the target speaker's speech characteristics and benefits the extracted speech quality. 

Historically, two prevailing strategies for cues extraction have been proposed in prior TSE works. One straightforward strategy is to use deep speaker embeddings as speaker cues, extracted from the enrollment utterance via pre-trained DNNs. These DNNs are well designed for speaker verification (SV) task, i.e., to testify whether the speaker identify of a test speech is the enrolled speaker. Those networks are typically trained over a large-scale dataset recorded in various acoustic conditions. As such, the learned deep speaker embeddings, as fixed-dimensional representations of variable-length utterances, are speaker-discriminative, meaning low intra-speaker variation and high inter-speaker discrepancy. Many deep speaker embeddings that are popular for SV are employed in recently proposed TSE models. Specifically, d-vector is applied in VoiceFilter \cite{wang19h}, VoiceFilter-lite \cite{wang20z_interspeech} and pDCCNR  \cite{eskimez22}; i-vector is used in a series of Speaekerbeam-based TSE models \cite{katerina19,chenglin19};  deep ResNet-based speaker embedding is used in an attention-based TSE \cite{li20p} and x-vector is adopted in recent dual-path filter network \cite{wang21x} .

Although deep speaker embeddings have been adopted in most TSE models, it is argued that they are separately optimized for SV and may be sub-optimal for TSE. Motivated by the opinion, another strategy is to extract cue from an auxiliary speaker encoder. Such an speaker encoder generally is jointly trained along with the TSE model by multi-task learning, using either multi-class classification loss (in \cite{katerina19,meng20,marc20}) or metric learning loss (in \cite{zifeng22}). 

Despite these efforts, the performance of the SOTA TSE model still does not outperform that of SS model. Indeed, on the widely adopted WSJ0-2mix dataset, the performance of the best TSE model is 18.6 dB, which is largely inferior to the state-of-the-art (SOTA) SS model (22.3 dB). This is contrary to the general expectation that the overall performance upper bound of TSE model is potentially higher than that of SS model, since the network only focuses on the reconstruction of the target speaker. 

We observe that the current speaker cue extraction approaches, whether originated from deep speaker embeddings or extracted from a jointly optimized speaker encoder, are designed with primary focus on the discriminability. However, it is well-known that speaker embedding features are scattered in high-dimensional feature space (generally higher than 128D). In the case, a classification separability, as a measure of discrimination effectiveness of embeddings, is unavoidably influenced by embedding variances within each class. Inspired by this, we hypothesize that speaker cue with clear class separability might be preferred for TSE. A simple yet novel idea that has not been explored yet in prior TSE studies.

To this end, as a preliminary investigation of our proposal, the scheme of sparse LDA-transform is explored in this paper. Specifically, we first build TSE baselines by integrating a SS model backbone with speaker cue of x-vector or a newly proposed speaker embedding, xi-vector (designed by incorporating uncertainty modeling in a speaker embedding neural network). Then we replace above speaker cues with their LDA-transformed versions, with the expectation that the latter that offer better classification power may facilitate the speaker extraction. Lastly, to bridge the performance gap between SOTA TSE and SS on the WSJ0-2mix, all our TSE models are built with the backbone of the SepFormer, one SOTA SS model.

The rest of the paper is organized as follows. Research works related to SepFormer and xi-Vector are briefed in Section 2. Section 3 describes our proposals. Experimental results are reported and analyzed in Section 4 and 5. Finally we conclude the paper in Section 6. 

\section{Related Works}
We start by giving a brief overview of the SepFormer \cite{subakan21}, the backbone of our proposed TSE models. SepFormer is a transformer-based single channel speech separation model. It is based on the learned-domain masking approach with overall architecture consisting of an encoder, a decoder, and a masking module. It processes the mixture input with the following pipeline: the encoder uses a convolution layer to transform the input into an STFT-like representation; then the masking module estimates masks from the representation; finally the decoder transforms the multiplications between the masks and representation into time-domain separated sources using a transposed convolution layer.

\begin{figure}
\includegraphics[width=\textwidth]{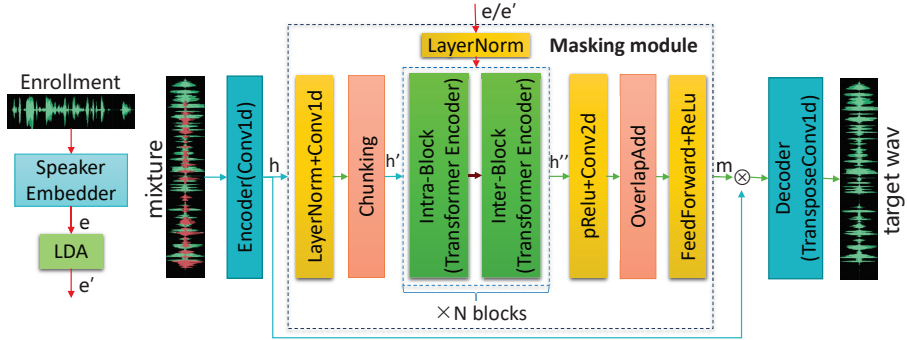}
\caption{Proposed TSE Architecture. Here $h$ refers to the learned speech features, $m$ denotes target source mask to mixture and $e$/$e'$ means raw speaker embedding and LDA transformed embedding.} \label{fig1}
\end{figure}

The core module of the pipeline is the masking module. It composes of transformers and follows the dual-path process to learn dependencies using a multi-scale approach. In particular, the mixture frames are firstly segmented into overlapping chunks. These stacked chunks are first processed independently along the chunks (intra-chunk processing) for short-term dependency and after that the dimensions are transposed and the sequence processing is executed across chunks (inter-chunk) to aggregate the information from all the chunks. This intra- and inter-chunk processes can be iteratively employed and repeated (by N times). The final aggregated information is processed by nonlinear activation following a linear layer and overlap-add operation to yield multiple masks (two masks in case of WSJ0-2mix). 

The SepFormer has shown impressive separation performance and achieved state-of-the-art accuracy in a list of benchmark datasets such as WSJ0-2mix\footnotemark[1] \footnotetext[1]{https://paperswithcode.com/sota/speech-separation-on-wsj0-2mix}and WSJ0-3mix\footnotemark[2], in terms of SI-SDRi  \footnotetext[2]{https://paperswithcode.com/sota/speech-separation-on-wsj0-3mix}. As a result, we choose it as the backbone for our TSE models proposed below. In addition, to boost the separation performance, we perform slight modification on the original SepFormer, by repeating the Intra-InterT processing from $N=2$ to $N=4$ times and simplifying the transformer encoder from 8 to 4 layers.



\section{Proposed Method}
In this section, we propose a few TSE models which are built by integrating different speaker cue with the separation backbone of SepFormer. These models share the same architecture as illustrated in Fig.\ref{fig1}. Differing from the SepFormer backbone, the TSE model introduces additional speaker cue $e/e'$ within the masking module and yields only one mask for the target speaker. More architecture details will be described later in our Experiment section.  

\subsection{TSE model with xi-vector}
As stated before, popular deep speaker embeddings, such as i-vector, d-vector, x-vector have been widely used as the speaker cue for TSE task. Recently, xi-vector embedding \cite{lee21} is proposed to incorporate frame uncertainty measures, that account for variability in the inputs, into deep speaker representation learning. In another words, it combines the benefit of generative nature of the i-vector with the speaker discrimination of the x-vector. 

Given the capability to take
into account frame-wise uncertainty, the proposed xi-vector
embeddings exhibit improve robustness to perturbation

By explicitly handling the uncertainty of input features, we expect the xi-vector can be more robust to perturbation than its predecessors (e.g. x-vector) thus is helpful for TSE. Thus, herein we propose to utilize the xi-vector as speaker cue, which has not been investigated yet in previous TSE works. 

\begin{figure}
\includegraphics[width=\textwidth]{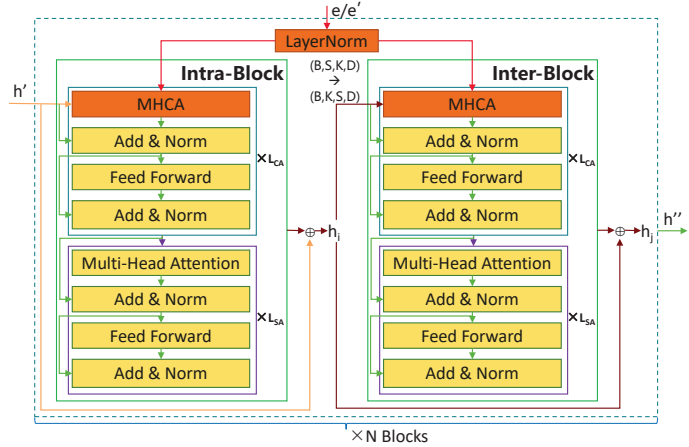}
\caption{ The masking module with adoption of our proposed speaker-speech fusion strategy. Here MHCA denotes Multi-Head Cross Attention. $B$=batchsize, $S$=number of chunks, $K$=frames in each chunk, $D$=speech feature dimension of each frame. $L_{CA}$ and $L_{SA}$ refer the layer numbers of transformer encoder with cross attention and self-attention, respectively. $\bigoplus$ denotes addition.} \label{fig2}
\end{figure}

To find proper approaches to leverage the new speaker cue in SepFormer, some preliminary experiments have been conducted. From the results, we empirically find that it is better to integrate the speaker cue with SepFormer at the 1st layer of both Intra- and Inter-Transformer block (as illustrated in Fig.\ref{fig2}). That means, xi-vector is repeatedly appended with our SepFormer (consisting of $N=4$ Intra-InterT processings) at 8 different positions. 

Meanwhile, within the masking module, multiple fusion configurations to effectively combine the speaker cue with inputs of Intra- or InterT block are explored and compared, such as concatenation-based, product-based, attention-based scaling, FiLM-based and cross-attention-based. Among them, the best performed cross-attention-based fusion technique \cite{wang21} is employed in our TSE models, where the speaker cue contributes to both attention weight and fusion bias. Such speaker-speech fusion process (as highlighted in orange in Fig.\ref{fig2}) can be expressed as:
    \begin{align*}
    A_{pc} & = f_{weight}(e,h',\theta_{Q_e,K_e,V_e},\theta_{Q_{h'},K_{h'},V_{h'}}) \\
    f_{pc} & = A_{pc}(e\theta_{V_e} + h\theta_{V_{h'}})
    \end{align*}

where $e$ denotes the speaker cue, $h'$ refers to the input of Intra- or InterT block, $\theta_{Q,K,V}$ are attention-parameters, $f_{weight}$ is the weights function, $A_{pc}$ is attention weights and $f_{pc}$ is the fused output via parallel concatenation. 

\subsection{TSE models with sparse LDA-transformed embeddings} 
As aforementioned, we hypothesize the speaker cue with high classification power might be beneficial to TSE. To verify it, a simple way is to perform LDA (Linear Discriminant Analysis) \cite{martinez01} on raw speaker embeddings. As a classic technique for dimensionality reduction, LDA projects original data into a lower-dimensional subspace (strictly less than the number of classes $M$), by focusing on maximizing the separability among the $M$ classes.

Generally, the class separability criteria are not easily computed. So in LDA, a Gaussian distribution is assumed. Under the assumption, optimal classes separation can be obtained to maximize the ratio of a between-class variance to a within-class variance via uncorrelated linear discriminants. 

Note that there is no loss in class separability power in reducing embedding dimension to $M-1$. While for the case of speaker cue for TSE, we think the separability and generalization property are both desired. Thus, we propose to adopt sparse LDA-transformed embeddings from a small subset (with $l$-dimension and $l<M-1$) spanned by the first $l$ discriminants. In another words, we only select important discriminants while eliminate other irrelevant or redundant ones by losing some amount of data variance. The yielded new speaker cues have the advantage of balanced separability and generalization by giving away a small portion of unwanted information and keeping as much separability as possible.


To apply the idea for the first time, we extend our TSE model that is proposed in the Section 3.1, by additionally transform the naive speaker cue (i.e. x-vector or xi-vector) via LDA prior to the speaker-speech fusion process.  

\section{Experimental Settings}
\subsection{Dataset}
\textbf{Dataset} WSJ0-2mix \cite{hershey16} is a popular dataset for speech separation, where single channel mixture of two speakers are created by randomly mixing utterances from the Wall Street Journal (WSJ0) corpus, at a sampling rate of 8 kHz. The utterances are mixed at various signal-to-noise (SNR) ratios between 0 dB and 5 dB. 

To investigate the effectiveness of our proposal, we build a TSE dataset based on the WSJ0-2mix, following the script\footnotemark[3] \footnotetext[3]{{https://github.com/xuchenglin28/speaker\_extraction\_SpEx/tree/master/data/\\wsj0\_2mix}}. In all, the generated TSE dataset contains training set (20000 utterances), validation set (5000 utterances) and test set (3000 utterances). All utterances in training and validation set are from 101 speakers, and utterances in test set are from 18 different speakers. The average duration of auxiliary speech is 7.3 s in test set.

To create dataset for speaker embeddings, we split official WSJ0 training set $si\_tr\_s$ into training set and validation set by splitting ratio of 0.95:0.05. The test set is constructed based on the combined WSJ0 development set $si\_dt\_05$ and evaluation set $si\_et\_05$. For each speaker present in the combined set , we randomly choose 200 utterances as positive speech samples and another 200 utterances from other speaker as negative speech samples.

\subsection{Implementation details}
Our proposed TSE system is implemented in PyTorch with architecture details listed below:
\begin{itemize}
\item Encoder module:

An 1D convolution layer is used in Encoder with a kernel size of 16 and a stride factor of 8, with input and output dimension of 1 and 256, respectively;

\item Masking module:

The Masking module mainly consist of chunking, overlap-add, Intra-Inter Transformer block and Conv2d. We set frames $K$=250 in each chunk, and set speech feature $D$=256. For each Intra- and Inter- Transformer Block, the layer of cross attention and self-attention is set to 1 and 3, respectively. Different with the SepFormer backbone, we use 256 output channel in Conv2d instead of 512 because only one speaker to extract for TSE task.

\item Decoder module:

As the inverse process of Encoder, Decoder module uses an 1D transposed convolution layer. It share the same kernel size and stride factor as the Encoder module, with input and output dimension of 256 and 1, respectively.
\end{itemize}
Given inputs of mixture and auxiliary speaker cue, our TSE system is trained under supervision of SI-SDR loss. Due to memory limitation, we choose batch size as 1. Our TSE systems are trained using Adam optimizer and following the learning rate scheduler of \textit{ReduceLROnPlateau}. The initial learning rate is set to $1.5e-4$ for the first 20 epochs then decreased to half of its previous setting when no improving loss on the validation is observed over 2 epochs. 


For LDA implementation, we employ \textit{sklearn} library implementation and adopt eigen decomposition as our solver. A few number of dimensions are set for the components parameter and leave all the rest as default. The library also provides the metric of \textit{explained variance ratio}, which accounts for how much variance each component explain in percentage.

\subsection{Evaluation metrics}
Follow the same evaluation method and metrics as adopted in most prior arts, our TSE systems are evaluated with reconstruction metrics of SDRi, and SI-SDRi and speech quality metric of PESQ. For them, higher value represents better performance. In this study, we report all metrics in order to make sufficient performance comparisons with more prior arts. In addition, speaker embeddings are evaluated using standard metrics of minimum detection cost function (minDCF) and equal-error-rate (EER). Generally speaking, the smaller the EER and minDCF, the better the discriminability performance of the embedding.  

\section{Results and Discussion}
\subsection{TSE Baseline}
Firstly, our speech separation backbone system is constructed by re-producing the codes provided by authors\footnotemark[4]\footnotetext[4]{{https://github.com/speechbrain/speechbrain/tree/develop/recipes/WSJ0Mix/separation}}. Our reproduced system is on par with the original SepFormer \cite{subakan21}, with slight performance degradation in terms of SI-SDRi. The claimed performance of original SepFormer with two blocks is 22.3 dB, our reproduced system achieves 21.6 dB and our modified SepFormer with four blocks further boosts performance to 21.9 dB. In the following, all our TSE systems are implemented based on backbone of the modified SepFormer. 

Based on the backbone, we first build our TSE baseline. It adopts the popular x-vector as speaker cue, so is termed as x-TSE system. Our x-vector model is trained following the SpeechBrain recipe\footnotemark[5]\footnotetext[5]{{https://github.com/speechbrain/speechbrain}}. For fair comparison with prior arts, the training set of x-vector is limited to the training and deviation data of WSJ0-2mix, which contains 101 speakers in total and entirely disjoints from the test set. 

Similarly, under the same training settings, our xi-vector model is trained by replacing the statistics pooling in x-vector with the Gaussian posterior inference. Only the posterior mean vector ($\phi$) is extracted for our xi-vector.

\begin{figure}
\includegraphics[width=\textwidth]{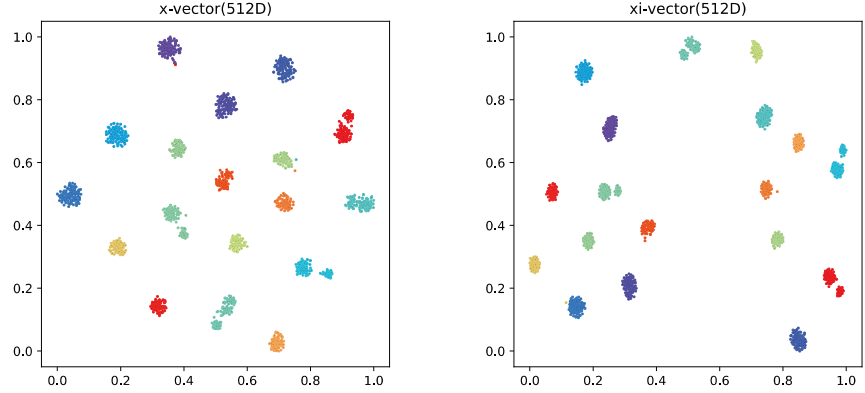}
\caption{Plot of embeddings visualized using t-SNE on our test set for speaker embeddings} \label{fig3}
\end{figure}

\begin{table}[]
    \centering
    \setlength{\tabcolsep}{5mm}{
    \begin{tabular}{cccc} \\ \hline
        speaker embedding  &  dimension & EER(\%) & minDCF \\ \hline
        x-vector & 512 & 3.58 & 0.364 \\ \hline 
        xi-vector & 512 & 3.31 & 0.370\\ \hline 
    \end{tabular}}
    \caption{Performance comparisons on our trained speaker embeddings}
    \label{tab:Tab_emb}
\end{table}
The corresponding performances of two embeddings are compared in Tab.\ref{tab:Tab_emb}. The comparison results show that xi-vector  gives moderate improvement on EER with marginal degradation on MinDCF. Also, by visualizing the 512D embeddings via t-SNE plots, Fig.\ref{fig3} indicates xi-vectors are indeed more clearly clustered than x-vector, although all the samples are spaced apart for both embeddings. 
In addition, it should be mentioned that the number of speakers in training data with WSJ0-2mix is 101, much smaller than the number in a standard speaker recognition task (e.g., over 1k in VoxCeleb1 and 6k in VoxCeleb2). 

\subsection{Proposed TSE Systems}
Except the baseline, a bunch of TSE systems are implemented following our proposals. They share the same TSE architecture as illustrated in Fig.\ref{fig1}, but with three forms of speaker cues: i) xi-vector; ii) LDA-transformed xi-vector and iii) LDA-transformed x-vector. Accordingly, these systems are named as xi-TSE, xi-LDA-TSE and x-LDA-TSE, respectively. 

\noindent\textbf{Proposed TSE systems vs. Baseline}
First, the proposed three TSE systems are compared to the baseline, x-TSE. In addition, considering the dimensional flexibility of LDA-transformation, for TSE systems adopting LDA-transformed embeddings, two different dimensionality settings are examined. Experimental results of all six TSE systems are listed in Tab.\ref{tab:vsbase}. Here, we underline the results of our baseline and bold-face the best results among all systems. In addition, the EER and minDCF are also reported as evaluation of the discriminability performance of associated speaker cues.  
\begin{table}[]
    \centering
    \begin{tabular}{c|c|c|c|c|c}
    \hline
    \hline
    & System & SI-SDRi(dB) & PESQ & minDCF & EER(\%) \\ \hline
    baseline &   x-TSE & \underline{17.1} & \underline{3.59} & \underline{0.36} & \underline{3.58}\\ \hline
    \multirowcell{5}[0pt][l] {our proposed \\ TSE system}   & x-LDA-TSE (64D) & 18.5 & 3.70 & 0.39 & 2.94\\ \cline{2-6}
       & x-LDA-TSE (32D) & 17.9 & 3.67 & \textbf{0.32} & 2.50 \\ \cline{2-6}
       & xi-TSE & 18.0 & 3.65 & 0.37 & 3.31 \\\cline{2-6}
       & xi-LDA-TSE (64D) & 18.2 & 3.67 & 0.43 & 2.75\\ \cline{2-6}
       & xi-LDA-TSE (32D) & \textbf{18.8} & \textbf{3.71} & 0.45 & 2.78\\ \hline\hline
    \end{tabular}
    \caption{Comparison of proposed TSE systems with the baseline on the WSJ0-2mix test set}
    \label{tab:vsbase}
\end{table}

From the results listed in Tab.\ref{tab:vsbase}, we have the following key observations and conclusions:
\renewcommand{\labelenumi}{\alph{enumi})}
\begin{enumerate}
    \item adopting LDA-transformation on baseline noticeably improves the baseline performance, which validates the effectiveness of our second scheme;
    \item a similar behavior is observed for xi-TSE, suggesting proposed LDA-transformation scheme might be general that can be applied to other speaker embeddings; 
    \item the performance of TSE system is sensitive to the dimensionality setting of LDA-transformation; optimal dimensionality may be decided empirically;
    \item xi-TSE outperforms the baseline in both SI-SDRi and PESQ, which meets our earlier expectation that the robustness of xi-vector is helpful for TSE;
    \item the TSE system with the minimum EER (minDCF) does not yield the best TSE performance. It again, indicates the potential objective mismatch between speaker verification and TSE.
\end{enumerate}

Observing the performance gains shown above, only xi-vector is further explored in subsequent experiments. 
\begin{table}[]
    \centering
    \begin{tabular}{c|l|c|c|c|c}
    \hline
    \hline
    & TSE System & training set& SDRi & SI-SDRi & PESQ \\ 
    & & of speaker cue & (dB) & (dB) & \\ \hline
    mixture & - & -& 0 & -0.001 & 2.01 \\ \hline
   \multirow{4}{*}{prior arts} &SpEx\cite{chenglin20} & \usym{2713}  & 16.3 & 15.8 & - \\ \cline{2-6}
    & Spex+\cite{meng20} & \usym{2713} & 17.2 & 16.9 & - \\ \cline{2-6}
    & Speakerbeam \cite{katerina19}& \usym{2713} & 9.7 & - & - \\ \cline{2-6}
    & Speakerbeam + DC \cite{katerina19}& \usym{2713} & 10.9 & - & - \\ \cline{2-6}
    & DPRNN-Spe-IRA\cite{luo20} &\usym{2717} & 17.6 & 17.3 & 3.43 \\ \cline{2-6}
    & WASE \cite{hao21} & \usym{2713} & 17.0 & - & - \\ \cline{2-6} 
    
    & SpExsc\cite{wang21} &\usym{2713} & 18.6 & 18.4 & - \\ \cline{2-6}
    & SpExpc\cite{wang21} &\usym{2713} & 18.8 & 18.6 & - \\
    \bottomrule
    \multirow{4}{*}{ours} & xi-TSE &\usym{2713} & 18.6 & 18.0 & 3.65 \\ \cline{2-6}
      & xi-LDA-TSE (32D) &\usym{2713} & 19.2 & 18.8 & 3.71\\ \cline{2-6}
    &  xi-TSE + DM  &\usym{2713} & 19.7 & 19.1 & 3.75\\
    \cline{2-6}
    & xi-LDA-TSE (32D) + DM  &\usym{2713} & \textbf{19.9} & \textbf{19.4} & \textbf{3.78}\\ \hline\hline
    \end{tabular}
    \caption{Comparison of proposed systems with the existing prior arts on the WSJ0-2mix test set}
    \label{tab:vsprior}
\end{table}

\noindent\textbf{Proposed TSE Systems vs. Prior Arts}
Now we compare our proposed TSE systems against eight top performing SOTA systems on the WSJ0-2mix. The overall comparison results are presented in Tab.\ref{tab:vsprior}. The upper part of the table presents the published results directly cited from literature and the last four rows correspond to experimental results from our proposed systems. 

Here $-$ means the corresponding result is not reported in the original paper. To have a fair comparison, the training data for speaker cue is also reported: the $\usym{2713}$ denotes the exact same dataset as our baseline model is used and $\usym{2717}$ indicates some large-scale dataset is employed.

From above results, we can observe that our xi-LDA-TSE (32D) system outperforms all existing methods, sometimes with a significant margin. In particular, to authors' best knowledge, the SpExpc is the most competitive existing work. Our xi-LDA-TSE (32D) system consistently surpasses it over all evaluation metrics. In addition, the use of dynamic mixing (DM) \cite{zeghidour21} data augmentation contributes significant performance improvement, by on-the-fly creating of new mixtures.  The last takeaway from Tab.\ref{tab:vsprior} is that our best performing system (with combined xi-LDA-TSE(32D) and DM) delivers the best TSE performance reported on the WSJ0-2mix.

\begin{figure}
\includegraphics[width=\textwidth]{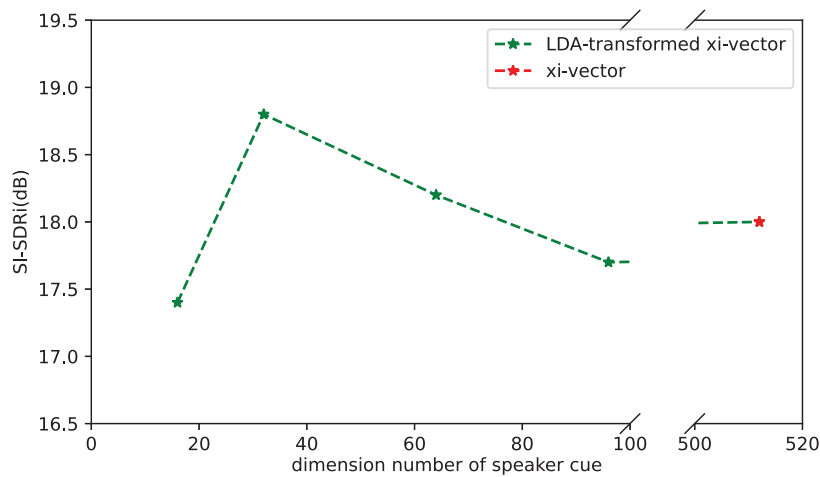}
\caption{Relationship between LDA dimensions of xi-vector and corresponding TSE performances in SI-SDRi} \label{fig4}
\end{figure}

\noindent\textbf{The optimal dimension of LDA-transformed speaker cue}
As presented in Tab.\ref{tab:vsbase}, the LDA dimensionality associated with speaker cue affects the performance of corresponding TSE system. So lastly, we investigate the contribution of dimensionality by testing more speaker cues with different dimensions. 
Fig.\ref{fig4} shows the value of SI-SDRi versus the number of dimensions defined by the LDA. From the figure, it can be observed that the optimal dimension of the xi-LDA based speaker cue is 32. We find that, from the metric \textit{explained variance ratio} provided by \textit{sklearn} library, such a speaker cue holds around 82\% of the total variation in xi-vector.   

\section{Conclusion}
In this study we investigate the speaker cue issue for target speaker extraction. In contrast with most existing TSE works that utilize speaker embedding and focus primarily on discriminability power, we propose to extract speaker cue with clear class separability. In particular, we adopt sparse LDA-transformed speaker embedding as new speaker cues. To verify the validity of our proposal, multiple TSE models are built based on the same architecture that differs only on associated speaker cue. Experimental results on benchmark dataset WSJ0-2mix validate the effectiveness of our proposals, showing significant performance improvements benefited from the LDA-transformed speaker cues. In addition, performance gains of our proposal are consistently found in TSE models with different speaker embeddings, which suggests that our proposal is independent of the embedding form. Further experiments on optimal LDA dimensionality reveal that proper speaker cues with balanced separability and generalization yield better TSE performance. In addition, our best TSE system outperforms existing SOTA TSE systems and achieves a new record of SI-SDRi and PESQ on the WSJ0-2mix. We believe that these empirical results are interesting and enlightening. Future research will evaluate our proposal on larger dataset for speech separation and further enhance it with the purpose of practical deployment.

\end{document}